\documentstyle[preprint,pra,aps]{revtex}

\begin{document}

\bibliographystyle{prsty}

\tighten

\title{Causality in quantum teleportation:\\
information extraction and noise effects\\ in 
entanglement distribution}
%on the transfer of effects from entanglement 
%distribution channels to the teleported state}

\author{Holger F. Hofmann}
\address{CREST, Japan Science and Technology Corporation (JST)\\ 
Research Institute for Electronic Science, Hokkaido 
University\\
Kita-12 Nishi-6, Kita-ku, Sapporo 060-0018, Japan}

\date{\today}

\maketitle

\begin{abstract}
Quantum teleportation is possible because entanglement allows
a definition of precise correlations between the non-commuting 
properties of a local system and corresponding non-commuting
properties of a remote system. In this paper, the exact causality
achieved by maximal entanglement is analyzed and the results are 
applied to the transfer of effects acting on the entanglement 
distribution channels to the teleported output state. In 
particular, it is shown how measurements performed on the 
entangled system distributed to the sender provide information 
on the teleported state while transferring the corresponding 
back-action to the teleported quantum state.
\end{abstract}

%\pacs{PACS numbers:
%03.65.Ud  % Entanglement and non-locality
%03.67.Hk  % Quantum communication
%03.65.Ta  % Foundations of quantum mechanics; measurement theory
%03.65.Yz  % Decoherence; open systems; quantum statistical methods
%}
\vspace{0.5cm}

\section{Introduction}
Quantum teleportation is one of the most fundamental 
applications of entanglement between well separated
physical systems \cite{Ben93,Bou97,Vai94,Bra98,Fur98}. 
The transmission
of a quantum state through local measurements and 
classical communications illustrates some of
the essential features of quantum physics.
It is therefore of great interest to analyze the 
transmission processes in detail, especially in 
realistic circumstances where entanglement may not
be maximal \cite{Mil99,Enk99,Opa00,Lee00,Hof00,Bra00,Bow01}. 
In this paper, the principles
of quantum teleportation are reviewed and a representation
of the teleportation process emphasizing the precise
causality implied by maximal entanglement is proposed.
This representation is especially useful to investigate
the transfer of effects on the entanglement distribution
channels to the teleported output state. The general theory
of this transfer is formulated and the results are applied
to measurements on the entangled signal sent from the source
of entanglement to the sender. Such measurements could be
used by a third party to extract information during the 
teleportation process, e.g. for the purpose of eavesdropping.
Moreover, the effects of this measurement illustrate the 
distribution of information and noise in the teleportation 
process, providing insights into the dynamics of quantum 
information processes. 

\section{Properties of entanglement}
\label{sec:entangle}
Entanglement is the quantum mechanical property that makes
teleportation possible. As was already pointed out by 
Schr\"odinger \cite{Sch35}, the essential feature of entanglement
is that all of the relative properties of two systems can be 
defined with precision if the individual properties are 
maximally uncertain.
In teleportation, the quantum state in the input can therefore
be reconstructed in the output by exclusively referring to 
these precise relations between the systems, while
avoiding any direct measurement of individual properties.

In order to express entanglement in terms of the relation
between physical properties in the two systems, it is 
useful to write the maximally entangled state in a common
basis $\mid \! n \rangle$. Maximal entanglement between two 
N-level systems, $R$ and $B$, can then be expressed by
the quantum state
\begin{equation}
\mid \! E_{\mbox{max.}}\rangle_{R,B} = \frac{1}{\sqrt{N}}
\sum_{n=0}^{N-1} \left(\hat{U}_0\mid \! n \rangle\right)_R
\otimes \mid \! n \rangle_B.
\end{equation}
The precise properties of this quantum state are defined by
the unitary transformation $\hat{U}_0$. Specifically, any
observable property $\hat{O}_B$ of system $B$ corresponds
to an observable $\hat{O}_R$ in system $R$, such that the 
measurement values obtained for $\hat{O}_B$ will always be
equal to the measurement values obtained for $\hat{O}_R$.
$\hat{U}_0$ defines the relation between $\hat{O}_B$
and $\hat{O}_R$ as
\begin{equation}
\label{eq:mirror}
\hat{O}_R = \hat{U}_0\hat{O}^T_B \hat{U}^{-1}_0,
\end{equation}
where the transpose $\hat{O}^T_B$ is defined with respect to 
the basis $\mid \! n \rangle$. 
It is then possible to formulate the unusual properties of
entanglement in the spirit of the Einstein-Podolsky-Rosen
paradox \cite{EPR,Rei89} as an apparent violation of local 
uncertainty relations by non-local correlations.
Given two non-commuting properties of system $B$, 
$\hat{X}_B$ and $\hat{Y}_B$, measurement results for both of
these two properties can be predicted by an observer in $R$
from measurements of the corresponding properties
$\hat{X}_R$ and $\hat{Y}_R$ since the maximally entangled 
state is an eigenstate of the two operator properties
\begin{eqnarray}
\hspace{1.5cm}
(\hat{X}_B-\hat{X}_R)\mid \! E_{\mbox{max.}}\rangle_{R,B} &=& 0
\nonumber \\ \mbox{and} \hspace{1cm}
(\hat{Y}_B-\hat{Y}_R)\mid \! E_{\mbox{max.}}\rangle_{R,B} &=& 0
\nonumber \\[0.3cm]
\mbox{with}\hspace{1cm} \hat{X}_R &=& \hat{U}_0\hat{X}^T_B \hat{U}^{-1}_0
\nonumber \\ 
\mbox{and}\hspace{1.2cm} \hat{Y}_R &=& \hat{U}_0\hat{Y}^T_B \hat{U}^{-1}_0.
\end{eqnarray}
Any measurement result of $\hat{X}_R$ in $R$ must be
equal to the measurement result for $\hat{X}_B$ in $B$, and 
any measurement result of $\hat{Y}_R$ in $R$ must be
equal to the measurement result for $\hat{Y}_B$ in $B$.
Interestingly, this property has a quite intuitive classical
interpretation. Effectively, the reference system $R$ is a
mirror image of system $B$. In classical physics, this property
is not unusual, since all properties of both systems can be
defined with precision: uncertainty is a specifically 
non-classical feature of quantum mechanics. Entanglement
is only difficult to understand because we cannot explain the
connection between the local uncertainty expressed by product
states and the non-local uncertainty expressed by inseparable
entangled states. 
Using the precisely defined relation between physical properties
in systems $R$ and $B$, it is therefore possible to explain 
quantum teleportation without using any purely quantum mechanical
terminology. 

\section{Principles of quantum teleportation}
\label{sec:tele}
In quantum teleportation, entanglement is applied to establish a
well-defined relation between an unknown input $A$ and an
output $B$ via the reference $R$. Initially, the output $B$
and the reference $R$ are maximally entangled. This means that
there exists a well-defined relation between the properties 
of $R$ and of $B$, while the individual properties of the
two systems are completely unknown. 
The relation between the unknown input $A$ and the reference
$R$ is then obtained from the joint measurement (also referred
to as a Bell measurement) of $A$ and $R$ at the location of the
sender. The sender communicates the information obtained
on the previously unknown relation between $A$ and $R$ to the
receiver, and the receiver obtains the exact relation of the
unknown input in $A$ and the quantum system $B$ by linking 
the measured relation between $A$ and $R$ with the previously 
known relation between $R$ and $B$. Figure \ref{physics} 
illustrates this analysis of teleportation. The letters $A$,
$R$, and $B$ denote a complete set of operator properties
of the respective systems, while $f_{\mbox{in}}$ and 
$f_m$ represent the well defined relations established by the
entanglement resource and the joint measurement. 

The main problem with this intuitive approach to quantum
teleportation is that it does not explicitly include the 
statistical properties of the quantum state. To express these
aspects of quantum teleportation, one should start with an 
input state $\mid \! \psi_{\mbox{in}}\rangle_{A}$ in $A$
and the entangled state $\mid \! E_{\mbox{max.}}\rangle_{R,B}$
in $R$ and $B$. This product state is then measured in the
subspace of $A$ and $R$. If a perfect Bell measurement is
performed, the systems are projected into a maximally 
entangled quantum state $\mid \! P(m)\rangle_{A,R}$, where
$m$ denotes the measurement result obtained. This maximally
entangled state can be conveniently expressed in terms of the 
$\hat{U}_0\mid \! n \rangle$ basis in $R$. It then reads
\begin{equation}
\mid \! P(m)\rangle_{A,R} =
\sqrt{\frac{\chi(m)}{N}}
\sum_{n=0}^{N-1} \left(\hat{U}(m)\mid \! n \rangle\right)_A
\otimes \left(\hat{U}_0\mid \! n \rangle\right)_R.
\end{equation}
The normalization factor $\chi(m)$ is necessary to provide the
correct probability for measuring $m$ if a complete set of
non-orthogonal measurements is considered.
The relation between properties in $R$ and properties in $A$
is expressed by the combination of the unitary 
transformations $\hat{U}_0$ and $\hat{U}(m)$. Since $\hat{U}_0$
represents the relation between $R$ and $B$, the unitary
transformation $\hat{U}(m)$
effectively describes the relation between $A$ and $B$ necessary
for the reconstruction of the quantum state.
This can be verified by applying the measurement projection to
the input state of the quantum teleportation.
\begin{equation}
\label{eq:grid}
\hspace{-1cm}
\begin{array}{lcccccc}
{\mbox{Initial state}} &
\frac{1}{\sqrt{N}}\sum_n & \mid \! \psi_{\mbox{in}} \rangle_A 
& \otimes 
& \hat{U}_0 \mid \! n \rangle_R & \otimes & 
\mid \! n \rangle_B \\[0.3cm]
{\mbox{Measurement}} &
\sqrt{\frac{\chi(m)}{N}} \sum_{n^\prime} & 
\left(\langle n^\prime \!\mid \hat{U}^{-1}(m)\right)_A 
& \otimes & \left(\langle n^\prime \!\mid \hat{U}_0^{-1}\right)_R
& & \\[0.3cm]
{\mbox{Conditional output}}\hspace{0.5cm}&
\frac{\sqrt{\chi(m)}}{N} \sum_n  &  
\multicolumn{3}{c}{\langle n \!\mid 
\hat{U}^{-1}(m)\mid \! \psi_{\mbox{in}}\rangle} 
& &  \mid \! n \rangle_B\\[0.4cm]
& \multicolumn{6}{c}{\hspace{-0.5cm} 
= \hat{U}^{-1}(m) \hspace{0.3cm}
\underbrace{
\frac{\sqrt{\chi(m)}}{N} 
\mid \! \psi_{\mbox{in}}\rangle_B
}_{ 
\mbox{Teleported state}}}
\end{array}
\end{equation}
The unitary transformation $\hat{U}(m)$ therefore describes
the relation between the unknown input in $A$ and the fluctuating
output $B$, even though there has been no interaction or other
connection between $A$ and $B$. To understand this effect, it
is useful to compare the calculation with the explanation of
teleportation illustrated in figure \ref{physics}. 
This analysis indicates that the measurement performed on
$A$ and $R$ merely obtains information about the previously
unknown state in $B$. This information can be represented by
a decomposition of the density matrix in $B$ into
subensembles corresponding to the different measurement results
$m$. The quantum mechanical features then arise from the 
non-classical properties of density matrix decompositions.
In the present case, it is possible to consider a complete 
Bell measurement with
\begin{equation}
\label{eq:complete}
\sum_m \mid P(m)\rangle \langle P(m) \mid_{A,R} = \hat{1}_{A,R}.
\end{equation}
If this condition is fulfilled, the initial density matrix
at $B$ can be decomposed into a mixture of unitary 
transformations of the input state  
$\mid \! \psi_{\mbox{in}}\rangle$ with
\begin {equation}
\label{eq:decom1}
\hat{\rho}_B = \sum_m \frac{\chi(m)}{N^2} 
\left(
\hat{U}^{-1}(m)\mid \! \psi_{\mbox{in}}\rangle
\langle \psi_{\mbox{in}} \! \mid\hat{U}(m)
\right)_B
 = 
\frac{1}{N} \hat{1}_B.
\end{equation}
Because of the properties of $\hat{U}(m)$ associated 
with the completeness relation (\ref{eq:complete}),
this decomposition is valid for any input state 
$\mid \! \psi_{\mbox{in}}\rangle$. 

By obtaining the measurement information $m$, 
the receiver identifies the subensemble of the density matrix 
according to its relation with the teleported state. 
This selection process can be understood entirely 
in analogy with classical physics. 
In quantum mechanics however, there is 
no fundamental decomposition of the density matrix 
simultaneously valid for all possible input states.
The selection of a subensemble density matrix therefore 
appears to be a choice between mutually exclusive possibilities. 
Interpretational problems arise if one tries to reconcile 
different decompositions with each other.
Quantum teleportation provides an input state dependent 
decomposition of the maximally mixed density matrix in $B$.
It is this dependence of the decomposition of the 
density matrix in $B$ on the quantum state in $A$
which appears to introduce a non-local effect beyond the 
classical non-locality of statistical correlations 
between remote objects. Nevertheless, each subensemble 
is always potentially contained in the initial mixed 
state of $B$, and an observer in $B$ will not be able 
to identify a specific decomposition without information
about the measurement performed on $R$. 

In the case of ideal quantum teleportation described in 
this section, no information whatsoever is obtained about 
the properties of the teleported state.
Both the information initially available and the information
obtained in the Bell measurement is about relations between 
the systems, none is about individual systems. However, 
this condition is only fulfilled if maximal entanglement is 
available. Any modification
to the entangled state of $R$ and $B$ changes the dynamics
of teleportation by modifying the information about the 
individual systems $R$ and $B$, as well as the implications 
of the Bell measurement for system $A$. In the following, 
the effects of such modifications will be investigated.

\section{Transfer of effects from entanglement distribution
to the output state}
The entanglement distribution channels may be subject to
a variety of effects, such as measurements, decoherence,
or unitary transformations \cite{Opa00,Hue01,Kni01}. 
Such effects can be described by operators 
$\hat{E}_R$ and $\hat{F}_B$ acting on the quantum states 
of these channels. $\hat{E}_R$ and $\hat{F}_B$ can represent
any combination of unitary operations and measurement 
projections. Decoherence effects can be represented by 
random mixtures of such operators. The general transfer 
properties derived in the following therefore apply to all
kinds of interactions with the entanglement distribution
channels. 

\begin{figure}
\setlength{\unitlength}{1.6pt}
\begin{center}
\begin{picture}(220,240)
%\put(0,0){\framebox(220,240){}}
\put(10,205){\makebox(60,20){
          $\mid \psi_{\mbox{in}}\rangle_{A}$}}
\put(38,200){\line(0,-1){23}}
\put(42,200){\line(0,-1){23}}
\put(40,170){\line(1,2){7}}
\put(40,170){\line(-1,2){7}}
\put(5,132){\framebox(70,36){}}
\put(5,150){\makebox(70,18){Measurement}}
\put(5,132){\makebox(70,18){Output $= m$}}

\put(52,109){\line(-1,2){8.5}}
\put(71,71){\line(-1,2){12}}
\put(49,109){\line(-1,2){8}}
\put(69,69){\line(-1,2){13}}
\put(40,130){\line(0,-1){12}}
\put(40,130){\line(4,-3){10}}

\put(39,92){\makebox(30,20){\large $\hat{E}_R$}}

\put(168,109){\line(1,2){8.5}}
\put(149,71){\line(1,2){12}}
\put(171,109){\line(1,2){8}}
\put(151,69){\line(1,2){13}}
\put(180,130){\line(0,-1){12}}
\put(180,130){\line(-4,-3){10}}

\put(151,92){\makebox(30,20){\large $\hat{F}_B$}}

\put(70,34){\framebox(80,36){}}
\put(70,52){\makebox(80,18){Source of}}
\put(70,34){\makebox(80,18){ Entanglement}}

\put(75,148){\line(1,0){80}}
\put(75,152){\line(1,0){80}}
\put(159,150){\line(-2,1){10}}
\put(159,150){\line(-2,-1){10}}

\put(85,165){\makebox(60,8){\small classical}}
\put(85,157){\makebox(60,8){\small communication}}

\put(162,133){\framebox(40,30){$\hat{U}(m)$}}

\put(178,170){\line(0,1){23}}
\put(182,170){\line(0,1){23}}
\put(180,200){\line(1,-2){7}}
\put(180,200){\line(-1,-2){7}}

\put(130,205){\makebox(100,20){$\hat{T}_{\mbox{out}}(m) 
          \mid \psi_{\mbox{in}}\rangle_{B}$}}
\end{picture}
\end{center}
\setlength{\unitlength}{1pt}

\caption{\label{operators} Illustration of the transfer of
effects from entanglement distribution channels to the
output state. The output effect $\hat{T}_{\mbox{out}}$ is
a function of the input effects $\hat{E}_R$, $\hat{F}_B$,
and of the measurement result $m$.}
\end{figure}
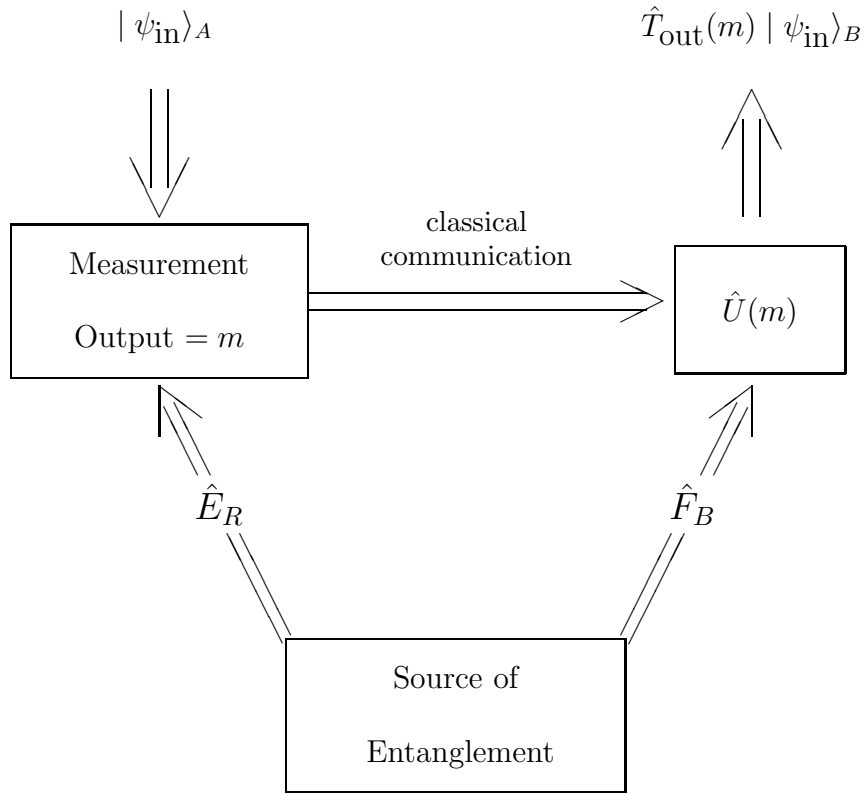

As shown in figure \ref{operators}, the effects on the 
entanglement distribution channels will be transferred to
a single effect on the output state, described by the 
output operator $\hat{T}_{\mbox{out}}$. This output operator
can be obtained by an analysis similar to the one applied
to the ideal teleportation case in equation (\ref{eq:grid}),
\begin{equation}
\label{eq:effects}
\hspace{-1cm}
\begin{array}{lcccccc}
{\mbox{Initial state}} &
\frac{1}{\sqrt{N}}\sum_n & \mid \! \psi_{\mbox{in}} \rangle_A 
& \otimes 
& \hat{E}_R \hat{U}_0 \mid \! n \rangle_R & \otimes & 
\hat{F}_B \mid \! n \rangle_B \\[0.3cm]
{\mbox{Measurement}} &
\sqrt{\frac{\chi(m)}{N}} \sum_{n^\prime} & 
\left(\langle n^\prime \!\mid \hat{U}^{-1}(m)\right)_A 
& \otimes & \left(\langle n^\prime \!\mid \hat{U}_0^{-1}\right)_R
& & \\[0.3cm]
{\mbox{Output}}\hspace{0.5cm}&
\frac{\sqrt{\chi(m)}}{N} \sum_{n,n^\prime}  &  
\langle n^\prime \!\mid 
\hat{U}^{-1}(m)\mid \! \psi_{\mbox{in}}\rangle &&
\langle n^\prime \!\mid 
\hat{U}_0^{-1}\hat{E}_R\hat{U}_0\mid \! n \rangle
& &  \hat{F}_B\mid \! n \rangle_B\\[0.4cm]
& \multicolumn{6}{c}{\hspace{-0.5cm} 
= \hat{U}^{-1}(m) \hspace{0.3cm}
\underbrace{
\frac{\sqrt{\chi(m)}}{N} \hat{U}(m)\hat{F}_B 
\left(\hat{U}_0^{-1}\hat{E}_R\hat{U}_0\right)^T 
\hat{U}^{-1}(m)
}_{\hat{T}_{\mbox{out}}}
\hspace{0.3cm}
\mid \! \psi_{\mbox{in}}\rangle_B .}
\end{array}
\end{equation}
The total effect on the teleported state can be separated 
into contributions from $\hat{F}_B$ and from $\hat{E}_R$.
The effect of $\hat{F}_B$ is only modified by the unitary 
transformation $\hat{U}(m)$, since this transformation
represents the only physical change of the output $B$ after 
the application of $\hat{F}_B$. 
The effect $\hat{E}_R$ on the reference channel $R$ 
is transferred to $B$ by the properties of entanglement 
since there is no direct physical interaction between $R$ and 
$B$. The contribution of $\hat{E}_R$ to $\hat{T}_{\mbox{out}}$ 
is equal to the operator $\hat{E}_B$ in $B$ corresponding 
to $\hat{E}_R$ according to equation (\ref{eq:mirror}). 
The effects on $R$ are thus transferred to $B$ by the precise 
correlations between the two systems. Effectively, any 
modification of the reference $R$ can be understood as a change 
in the relation between $R$ and $B$. Since nothing is known 
about the individual systems, it does not matter whether the 
effect really acts on $R$ or on $B$. 
The effect on the teleported state can then 
be written as a sequence of effects on $B$ transformed 
by $\hat{U}(m)$, 
\begin{equation}
\label{eq:Trans}
\hat{T}_{\mbox{out}} = \frac{\sqrt{\chi(m)}}{N}
\hat{U}(m) \hat{F}_B \underbrace{
\left(\hat{U}_0^{-1}\hat{E}_R\hat{U}_0\right)^T}_{\hat{E}_B} 
\hat{U}(m)
\end{equation}
It may be interesting to note that $\hat{E}_B$ acts before
$\hat{F}_B$, indicating that entanglement always connects
the past of system $B$ to system $R$, regardless of the 
actual sequence of $\hat{E}_R$ and $\hat{F}_B$ in time.

The mathematical properties of the formalism clearly show
that actions on $R$ and actions on $B$ have equivalent effects.
However, the entanglement distribution channels are usually 
well separated in space, and sometimes even in time. It is
therefore interesting to trace the causality connecting
actions on $R$ to the output in $B$ in more detail. Since the
most important aspect of quantum teleportation is the distribution
of information, it is convenient to focus this discussion 
on the possibility of extracting information in the 
reference channel $R$ by minimal back-action measurements.

\section{Measurements on an entanglement distribution channel}
In ideal quantum teleportation, no information about the
teleported quantum state is obtained in the process.
Since any quantum measurement has a corresponding back-action,
changing the original state unless it happens to be an
eigenstate of the measurement operator, this lack of information
about the teleported state is a necessary requirement for
the precise transfer of any unknown quantum state.
As discussed in section \ref{sec:tele}, this requirement is 
fulfilled because neither the original entanglement nor the
Bell measurement reveals any information on the individual 
systems. Instead, there is a perfect connection of precise
relations between the systems defined by entanglement 
properties.
However, this perfect connection can be broken
at any point. For example, an eavesdropper might decide to
"listen in" on the teleportation by tapping the entanglement
distribution line for the reference $R$. Since the density 
matrix of $R$ is completely random, the information initially
obtained is pure noise. However, the measurement result provides
information on the reference used in the Bell measurement. By
combining the measurement information $m$ with the noisy result
from the entanglement distribution line, information on the 
teleported quantum state is obtained. At the same time, the 
measurement back-action has changed the relation between $R$
and $B$, making it impossible to reconstruct the exact input
state. The output state is therefore modified by a measurement
back-action equal to the effects of a direct measurement performed
on the input state.

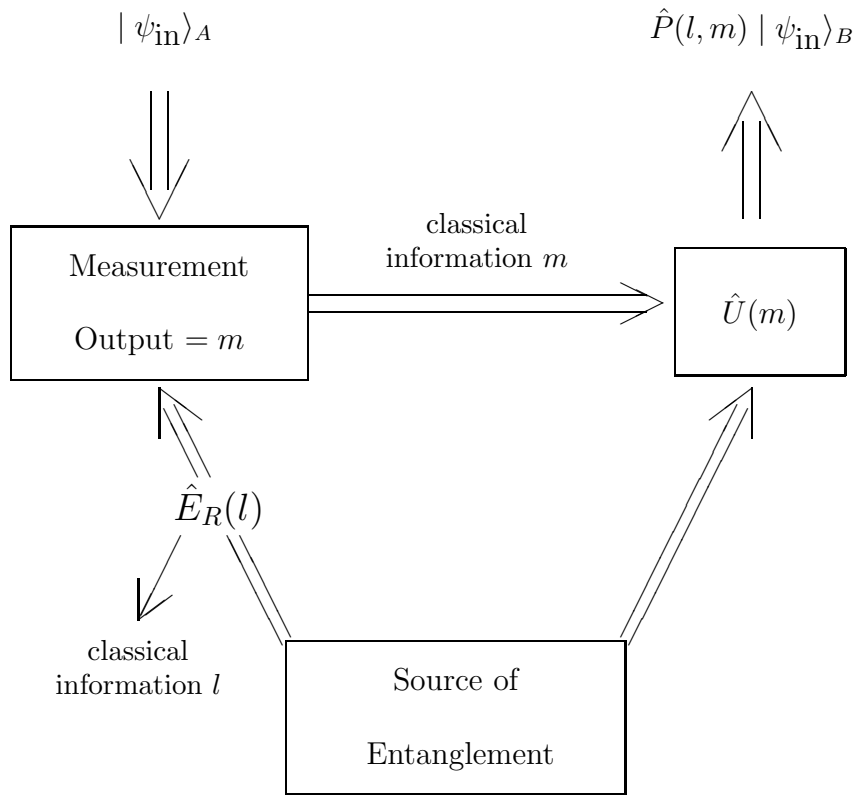
\begin{figure}
\setlength{\unitlength}{1.6pt}
\begin{center}
\begin{picture}(220,240)
%\put(0,0){\framebox(220,240){}}
\put(10,205){\makebox(60,20){
          $\mid \psi_{\mbox{in}}\rangle_{A}$}}
\put(38,200){\line(0,-1){23}}
\put(42,200){\line(0,-1){23}}
\put(40,170){\line(1,2){7}}
\put(40,170){\line(-1,2){7}}
\put(5,132){\framebox(70,36){}}
\put(5,150){\makebox(70,18){Measurement}}
\put(5,132){\makebox(70,18){Output $= m$}}

\put(52,109){\line(-1,2){8.5}}
\put(71,71){\line(-1,2){12}}
\put(49,109){\line(-1,2){8}}
\put(69,69){\line(-1,2){13}}
\put(40,130){\line(0,-1){12}}
\put(40,130){\line(4,-3){10}}

\put(39,92){\makebox(30,20){\large $\hat{E}_R(l)$}}

\put(45,95){\line(-1,-2){10}}
\put(35,75){\line(4,3){8}}
\put(35,75){\line(0,1){10}}
\put(20,64){\makebox(30,8){\small classical}}
\put(20,56){\makebox(30,8){\small information $l$}}

\put(149,71){\line(1,2){27.5}}
\put(151,69){\line(1,2){28}}
\put(180,130){\line(0,-1){12}}
\put(180,130){\line(-4,-3){10}}

\put(70,34){\framebox(80,36){}}
\put(70,52){\makebox(80,18){Source of}}
\put(70,34){\makebox(80,18){ Entanglement}}

\put(75,148){\line(1,0){80}}
\put(75,152){\line(1,0){80}}
\put(159,150){\line(-2,1){10}}
\put(159,150){\line(-2,-1){10}}

\put(85,165){\makebox(60,8){\small classical}}
\put(85,157){\makebox(60,8){\small information $m$}}

\put(162,133){\framebox(40,30){$\hat{U}(m)$}}

\put(178,170){\line(0,1){23}}
\put(182,170){\line(0,1){23}}
\put(180,200){\line(1,-2){7}}
\put(180,200){\line(-1,-2){7}}

\put(130,205){\makebox(100,20){$\hat{P}(l,m)
          \mid \psi_{\mbox{in}}\rangle_{B}$}}
\end{picture}
\end{center}
\setlength{\unitlength}{1pt}

\caption{\label{tap} Measurement on the reference channel
of entanglement distribution. $\hat{P}(l,m)$ is the effective
measurement operator representing both the information about
the quantum state and the measurement back-action associated
with the combined measurement results $l$ and $m$.
}
\end{figure}

Figure \ref{tap} illustrates this eavesdropping scheme.
The measurement performed on the entanglement distribution
channel is represented by the self-adjoint
operators $\hat{E}_R(l)$ corresponding to the measurement 
results $l$ of a minimal back-action measurement \cite{Wis95}.
According to equation (\ref{eq:Trans}), the effect of this 
measurement on the 
output state of the teleportation is described by the output
operator
\begin{equation}
\label{eq:Pop}
\hat{P}(l,m) = \frac{\sqrt{\chi(m)}}{N}
\hat{U}(m) \left(\hat{U}_0^{-1}\hat{E}_R(l)\hat{U}_0\right)^T
\hat{U}^{-1}(m).
\end{equation}
This operator is also self-adjoint, indicating that $\hat{P}(l,m)$
also represents a minimal back-action measurement.
Effectively, the operator acting on the teleported state is
a unitary transformation of the original measurement operator
$\hat{E}_R(l)$. The measurement performed on the entanglement
distribution channel $R$ therefore converts the teleportation
process into a measurement of the unknown input state.
The information obtained in this measurement is described
by the dependence of the probability of the measurement 
results $l$ and $m$ on the input state,
\begin{eqnarray}
\label{eq:leak}
p(l,m) &=& \langle \psi_{\mbox{in}} \! \mid
\hat{P}^2(l,m)
\mid \! \psi_{\mbox{in}}\rangle
\nonumber \\ 
&=& \frac{\chi(m)}{N^2}
\langle \psi_{\mbox{in}} \! \mid
\hat{U}(m) \left(\hat{U}_0^{-1}\hat{E}^2_R(l)\hat{U}_0\right)^T
\hat{U}^{-1}(m)
\mid \! \psi_{\mbox{in}}\rangle.
\end{eqnarray}
Equation (\ref{eq:leak}) describes the information extraction
achieved by the measurement on the entanglement distribution
channel $R$ by combining the information $l$ obtained from
$R$ with the information $m$ obtained in the Bell measurement.
Specifically, $\hat{P}^2(l,m)$ is the positive operator valued 
measure of the eavesdropping process, and different input states 
may be distinguished by the eavesdropper according to the 
relative statistical weight assigned to them by this measure 
\cite{Nie}.

Equation (\ref{eq:Pop}) shows that the back-action on the 
teleported state is minimal. It is therefore possible to 
realize an optimal eavesdropping scheme by performing minimal
back-action measurements on the entanglement distribution 
channels. The precise effect of eavesdropping on the teleported 
state is given by
\begin{equation}
\label{eq:backact}
\mid \! \psi_{\mbox{out}}\rangle
= \frac{1}{\sqrt{p(l,m)}} 
\hat{P}(l,m) \mid \! \psi_{\mbox{in}}\rangle.
\end{equation}
The eavesdropping attempt thus modifies the output state,
reducing the fidelity of quantum teleportation. A quantitative
expression for this loss of fidelity is given by the
overlap between the input state and the output state,
\begin{equation}
\label{eq:condF}
F(l,m) = |\langle \psi_{\mbox{out}} \! \mid
\! \psi_{\mbox{in}}\rangle|^2 = 
\frac{|\langle \psi_{\mbox{in}} \! \mid 
\hat{P}(l,m)
\mid \! \psi_{\mbox{in}}\rangle|^2}{p(l,m)}. 
\end{equation}
In general, this overlap may strongly depend on the measurement 
outcome. 
Equation (\ref{eq:condF}) is therefore a conditional fidelity
\cite{Hof00}. If the measurement results are not known, the 
total fidelity is given by the average over all possible outcomes, 
\begin{equation}
\label{eq:Ftot}
F_{\mbox{total}} = \sum_{l,m}|\langle \psi_{\mbox{in}} \! \mid
\hat{P}(l,m)
\mid \! \psi_{\mbox{in}}\rangle|^2.
\end{equation}
Thus the measurement operators $\hat{P}(l,m)$ given in equation 
(\ref{eq:Pop}) fully characterize the loss of fidelity
due to the eavesdropping attempt. 

Two particularly interesting features of this scheme are 
the distribution of information about the teleported state 
between the two measurement results $m$ and $l$, and 
the origin of the measurement back-action on the teleported
quantum state from a lack of information concerning the 
the actual relation between the input state in $A$ and 
the output $B$.
In the next section, these features will be analyzed in 
greater detail.

\section{Distribution of information and noise}
Equation (\ref{eq:leak}) describes the information extracted
from the teleported state in terms of the joint probability
$p(l,m)$ of obtaining a measurement result of $l$ in $R$,
followed by a Bell measurement result of $m$. The individual
probabilities for measuring $l$ and $m$ can be determined from
this joint probability by 
\begin{eqnarray}
p(l) &=& \sum_m p(l,m)
\nonumber \\ \mbox{and} \hspace{0.5cm}
p(m) &=& \sum_l p(l,m).
\end{eqnarray}

It is obvious that $p(l)$ should not depend on the input
state, since there is no relation between the entangled
state in $R,B$ and the input state in $A$ before the 
Bell measurement is performed. This independence of $p(l)$
from input state properties can be verified by applying the
completeness relation in equation (\ref{eq:decom1}) to
the sum over $m$,
\begin{eqnarray}
\label{eq:pl}
p(l) &=& \sum_m \frac{\chi(m)}{N^2}
\langle \psi_{\mbox{in}} \! \mid
\hat{U}(m) \left(\hat{U}_0^{-1}\hat{E}^2_R(l)\hat{U}_0\right)^T
\hat{U}^{-1}(m)
\mid \! \psi_{\mbox{in}}\rangle
\nonumber \\ &=&
\mbox{Tr} \left\{ 
\left(\hat{U}_0^{-1}\hat{E}^2_R(l)\hat{U}_0\right)^T
\left(\sum_m \frac{\chi(m)}{N^2} 
\hat{U}^{-1}(m)
\mid \! \psi_{\mbox{in}}\rangle
\langle \psi_{\mbox{in}} \! \mid\hat{U}(m) \right)\right\}
\nonumber \\
&=& \frac{1}{N} \mbox{Tr}\{\hat{E}^2_R(l) \}.
\end{eqnarray}
Note that this probability may also be derived directly from
the local density matrix in $R$ before the measurements.
This density matrix may also be expressed in terms of
the mixture given for $\hat{\rho}_B$ in equation (\ref{eq:decom1}).
Therefore, the sum over $m$ restores the situation before the
Bell measurement in $R$, providing the input independent 
probability $p(l)$. 

The probability $p(m)$ can be derived by making use of the
completeness relation of the minimal back-action measurement
of $l$,
\begin{equation}
\sum_l \hat{E}^2_R(l) = \hat{1}_R.
\end{equation}
In terms of the measurement back-action on the quantum state
in $R$, this completeness relation implies that the density 
matrix in $R$ after the measurement of $l$ is still equal 
to $\hat{1}_R/N$ if the measurement result is unknown. 
Therefore, the statistics of the Bell measurement is unchanged
and the probability $p(m)$ does not depend on the input state
either. In accordance with this observation, the result of the 
summation over $l$ reads
\begin{equation}
\label{eq:pm}
p(m) = \frac{\chi(m)}{N^2}
\langle \psi_{\mbox{in}} \! \mid
\hat{U}(m) \left(\hat{U}_0^{-1}
\left(
\sum_l \hat{E}^2_R(l)
\right)
\hat{U}_0\right)^T
\hat{U}^{-1}(m)
\mid \! \psi_{\mbox{in}}\rangle
= \frac{\chi(m)}{N^2}.
\end{equation}
Since the measurement in $R$ does not change the overall
statistics of the Bell measurement, an eavesdropping attempt 
using a minimal back-action measurement on $R$
cannot be detected by the sender, even if some statistical 
properties of the input states are known. By itself, the
measurement result $m$ does not provide any information on 
the input state. Such information is only obtained by 
combining the measurement results $m$ with the measurement
results $l$ from the entanglement distribution channel $R$.

As shown in equation (\ref{eq:decom1}), the main effect of
the Bell measurement is the decomposition of the maximally
mixed density matrix in $B$ in terms of unitary transformations
of the unknown input state. For a maximally entangled state,
this decomposition is equally possible for any input state.
However, the measurement of $l$ in $R$ provides information
about the decomposition of $\hat{\rho}_B$ which is independent of
the input state. It is therefore interesting to analyze the
decomposition of $\hat{\rho}_B$ in the presence of the measurement
on $R$. The total decomposition can be obtained from
$\hat{P}(l,m)$ if the conditional unitary transformation 
$\hat{U}(m)$ of the output is reversed. It then reads
\begin{equation}
\label{eq:decom2}
\hat{\rho}_B = \sum_{l,m} 
\hat{U}^{-1}(m) \hat{P}(l,m)
\mid \! \psi_{\mbox{in}}\rangle
\langle \psi_{\mbox{in}} \! \mid
\hat{P}(l,m) \hat{U}(m)
= \frac{1}{N} \hat{1}_B
.
\end{equation}
The decomposition into contributions 
with different $l$ represents the information obtained through
the measurement of $R$, while the teleportation effects are
represented by the decomposition into different $m$. 
This sequence of the decomposition can be expressed by
writing $\hat{\rho}_B$ as
\begin{equation}
\hat{\rho}_B = \sum_{l} 
\left(\hat{U}_0^{-1}\hat{E}_R(l)\hat{U}_0\right)^T
\underbrace{\left(\sum_m \frac{\chi(m)}{N^2} 
\hat{U}^{-1}(m)
\mid \! \psi_{\mbox{in}}\rangle
\langle \psi_{\mbox{in}} \! \mid\hat{U}(m) 
\right)_B}_{= \hat{1}_B/N}
\left(\hat{U}_0^{-1}\hat{E}_R(l)\hat{U}_0\right)^T
\end{equation}
Thus the measurement of $l$ first decomposes the density 
matrix in $B$ according to the information obtained from $R$
only. The measurement of $m$ then decomposes the components
of each result $l$ according to the same statistical weights
previously obtained for ideal teleportation in equation
(\ref{eq:decom1}). However, this decomposition is now modified
according to the measurement information provided by $l$,
resulting in a distortion of the transformed input state
components $\hat{U}^{-1}(m)\mid \! \psi_{\mbox{in}}\rangle$.
This distortion reflects the reduction of entanglement caused
by obtaining information on the local systems $R$ and $B$. 
The local uncertainty introduced into $R$ by the measurement
back-action is thus transferred to the output state.

It is possible to vary the measurement operators $\hat{E}_R(l)$
continuously between the unit operator $\hat{1}_R$ representing
no interaction and precise projections onto a complete 
orthonormal set of eigenstates $\mid \! l \rangle$ of a 
self-adjoint operator $\hat{L}_R$. In the latter case, the
entanglement is completely removed by the measurement. 
It is therefore a particularly simple example and may help to
illustrate some of the general features of the information
distribution caused by the measurement in $R$. 
The projective measurement in $R$ decomposes the density
matrix in $B$ into eigenstates $\mid \! \phi_l \rangle$
of the variable $\hat{L}_B$ corresponding to $\hat{L}_R$
according to (\ref{eq:mirror}). The measurement in $m$ 
cannot subdivide this decomposition any more, so it merely
provides measurement probabilities,
\begin{eqnarray}
\hat{\rho}_B &=& 
\sum_l \mid \! \phi_l \rangle \langle \phi_l\! \mid 
\sum_m p(l,m),
\nonumber \\ \mbox{with} &&
p(l,m)=
\frac{\chi(m)}{N^2}
|\langle \phi_l \mid \hat{U}^{-1}(m) 
        \mid \! \psi_{\mbox{in}}\rangle|^2.
\end{eqnarray}
As the measurement probability $p(l,m)$ shows, the Bell 
measurement now projects the input states onto unitary
transforms $\hat{U}(m)  \mid \! \phi_l \rangle $ of 
the eigenstates of $\hat{L}_B$. 
This corresponds to a precise measurement of
the property $\hat{U}(m)\hat{L}_B\hat{U}^{-1}(m)$ in
$A$. Only this property can then be reproduced in the
output state. The measurement back-action
randomizes the relation between properties that do not
commute with $\hat{L}_R$ in $R$ and their corresponding 
properties in $B$. Therefore, the Bell measurement does
not provide any information on such variables.
This simple example also illustrates the role of $l$
and $m$ in defining the effective measurement performed
on the teleported state. Knowledge of $m$ determines the 
actual variable $\hat{U}(m)\hat{L}_B\hat{U}^{-1}(m)$ 
defined by the measurement of $m$, while $l$ provides 
the measurement outcome for that variable. The measurement
result $l$ initially provides only information about a 
physical property of the reference system $R$, and the 
measurement result $m$ is necessary to establish the 
relation between this property in $R$ and a corresponding 
property of the input $A$. Thus, the Bell measurement 
randomly selects the physical property 
$\hat{U}(m)\hat{L}_B\hat{U}^{-1}(m)$ measured in $A$ 
after the measurement result $l$ has been obtained in $R$.

\section{Conclusions}
Quantum teleportation is an application of the extremely
precise correlations possible between two entangled systems.
The discussion presented in this paper shows how these
correlations can be identified in the quantum mechanical
formalism using unitary operations and density matrix
decompositions. The resulting formulation is especially
convenient for tracing the transfer of effects from
entanglement distribution channels to the output state.

In the case of a measurement on an entanglement distribution
channel, the measurement back-action introduces noise
into the teleportation by reducing the precision in the
correlation between the entangled systems. The information
obtained in the measurement distribution channel may then be
combined with the result of the Bell measurement to provide
information about the teleported quantum state. This example
thus illustrates how the quantum information in the original
input state is distributed between the two measurement results 
and the teleported quantum state in the output.

%============================================================

\begin{figure}
\setlength{\unitlength}{1.6pt}
\begin{center}
\begin{picture}(220,240)
%\put(0,0){\framebox(220,240){}}
\put(10,205){\makebox(60,20){Input $A$}}
\put(38,200){\line(0,-1){23}}
\put(42,200){\line(0,-1){23}}
\put(40,170){\line(1,2){7}}
\put(40,170){\line(-1,2){7}}
\put(5,132){\framebox(70,36){}}
\put(5,150){\makebox(70,18){Measurement}}
\put(5,134){\makebox(70,16){$A=f_m(R)$}}

\put(52,109){\line(-1,2){8.5}}
\put(71,71){\line(-1,2){12}}
\put(49,109){\line(-1,2){8}}
\put(69,69){\line(-1,2){13}}
\put(40,130){\line(0,-1){12}}
\put(40,130){\line(4,-3){10}}

\put(39,92){\makebox(30,20){Reference $R$}}

\put(168,109){\line(1,2){8.5}}
\put(149,71){\line(1,2){12}}
\put(171,109){\line(1,2){8}}
\put(151,69){\line(1,2){13}}
\put(180,130){\line(0,-1){12}}
\put(180,130){\line(-4,-3){10}}

\put(151,92){\makebox(30,20){Output $B$}}

\put(70,34){\framebox(80,36){}}
\put(70,52){\makebox(80,18){Entanglement}}
\put(70,34){\makebox(80,18){$R=f_{\mbox{in}}(B)$}}

\put(75,148){\line(1,0){60}}
\put(75,152){\line(1,0){60}}
\put(139,150){\line(-2,1){10}}
\put(139,150){\line(-2,-1){10}}

\put(75,157){\makebox(60,8){\small Information $m$}}

\put(142,132){\framebox(70,36){}}
\put(142,150){\makebox(70,18){Output relation:}}
\put(142,134){\makebox(70,16){
$A=f_m\left(f_{\mbox{in}}(B)\right)$}}

\end{picture}
\end{center}
\setlength{\unitlength}{1pt}

\caption{\label{physics} Illustration of causality in the
quantum teleportation process. The letters $A$, $R$, $B$ denote
the physical properties of the systems. The initial 
entanglement is described by the relation $f_{\mbox{in}}$ and
the result of the Bell measurement is given by $m$, corresponding
to a relation $f_m$ between the input $A$ and the reference $R$.}
\end{figure}
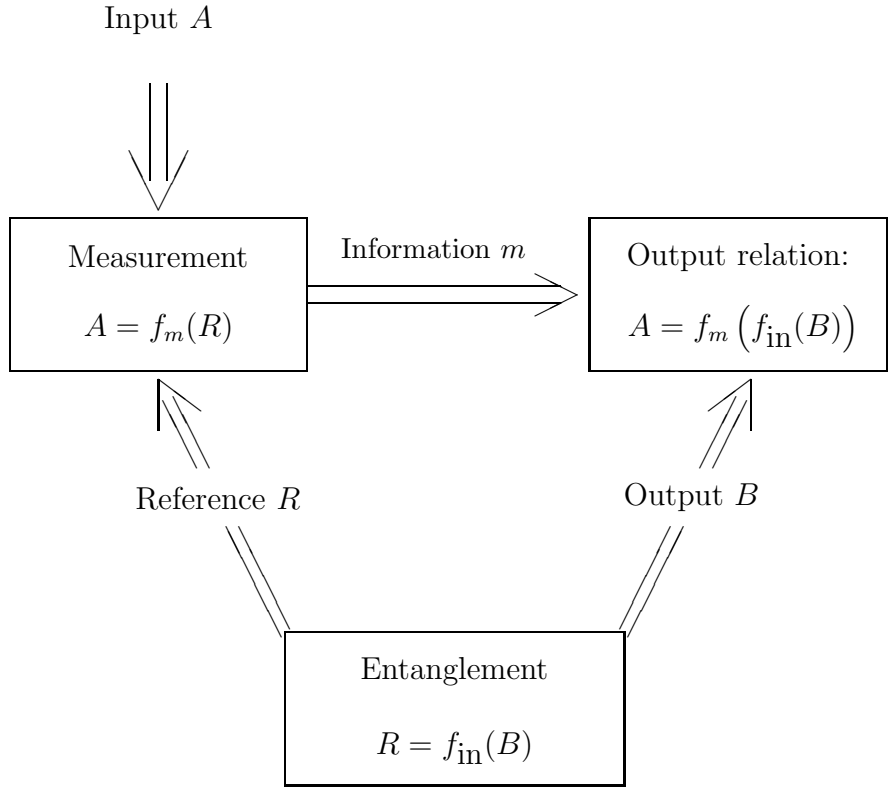

%========================================================

%========================================================

\end{document}